# Virtual Memory Streaming Technique for Virtual Machines (VMs) for Rapid Scaling and High Performance in Cloud Environment

A B M Moniruzzaman, *Student Member,IEEE,*
Kawser Wazed Nafi,
Syed Akther Hossain, *Member, IEEE & ACM*

*Abstract*— This paper addresses the impact of Virtual Memory Streaming (VMS) technique in provisioning virtual machines (VMs) in cloud environment. VMS is a scaling virtualization technology that allows different virtual machines rapid scale, high performance, and increase hardware utilization. Traditional hypervisors do not support true no-downtime live migration, and its lack of memory oversubscription can hurt the economics of a private cloud deployment by limiting the number of VMs on each host. VMS brings together several advanced hypervisor memory management techniques including granular page sharing, dynamic memory footprint management, live migration, read caching, and a unique virtual machine cloning capability. An architecture model is described, together with a proof-of-concept implementation, that VMS dynamically scaling of virtualized infrastructure with true live migration and cloning of VMs. This paper argues that VMS for Cloud allows requiring significantly reduced server memory and reducing the time for virtualized resource scaling by instantly adding more virtual machines.

## I. INTRODUCTION

Virtual memory [1] is a well-known technique used in most general-purpose operating systems, and almost all modern processors have hardware to support it. Virtual memory creates a uniform virtual address space for applications and allows the operating system and hardware to handle the address translation between the virtual address space and the physical address space [2]. Modern computers are sufficiently powerful to use virtualization to present the illusion of many smaller virtual machines (VMs), each running a separate operating system instance [3].

When running a virtual machine, the hypervisor creates a contiguous addressable memory space for the virtual machine. This memory space has the same properties as the virtual address space presented to the applications by the guest operating system. This allows the hypervisor to run multiple virtual machines simultaneously while protecting the memory of each virtual machine from being accessed by others. Therefore, from the view of the application running inside the virtual machine, the hypervisor adds an extra level of address translation that maps the guest physical address to the host physical address. As a result, there are three virtual memory layers in typical hypervisors: guest virtual memory, guest physical memory, and host physical memory.

Virtualization is a fundamental component in cloud technology. The VMM's provide Virtual Machine Memory by shadow paging and control with the help of guest operating system with mapping in the shadow page table [4], [5], [6] and reclaim virtual memory as required [5].

After the introduction, the remainder of the paper is structured as follows. First of all, section 2 describes the Virtual Memory Streaming Technique. Then, section 3 describes the proposed architecture and implementation details of the solution developed to enable virtual memory streaming for VMs in order to automatically scaling of virtual infrastructure. Next, section 4 describes a case study that executes a synthetic application that reproduces several dynamic memory consumption patterns on a virtual infrastructure. Later, section 5 discusses the main implications of the results both from the point of view of the user and the resource provider. Finally, section 6 summarizes the paper and points to future work.

## II. VIRTUAL MEMORY STREAMING TECHNIQUE

Virtual Memory Streaming (VMS) is a scaling virtualization technology that allows different virtual machines rapid scale, high performance, and increase hardware utilization. VMS brings together several advanced hypervisor memory management techniques including granular page sharing, dynamic memory footprint management, live migration, cloning, read caching, and a unique virtual machine cloning capability.

With VMS, virtual machines do not have to go through a performance intensive boot process. Instead, they launched from pre-booted live images and are ready to serve requests within seconds. The launched virtual machines have a low memory footprint, resulting in dense memory oversubscription. Using VMS, you can create a live image of a running virtual machine instance. The entire disk and memory state of the running instance including the operating system and applications is preserved in the live image. VMS optimizes entire virtualization layer providing faster boot times for virtual machines, generous runtime memory savings and a superior scale-out platform for application.

VMS technology takes a snapshot of running VM of the Operating System loaded and application ready to service required. In the figure 2.1 shows, snapshot is combination of OS-memory; application memory and runtime data.

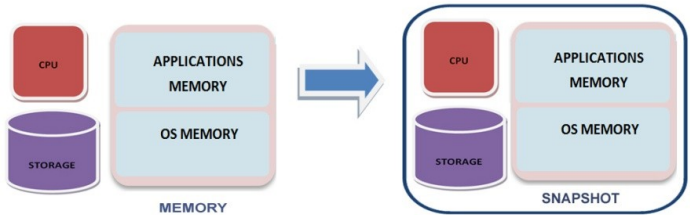

Figure 2.1: Snapshot

From this snapshot, this provision clones can start instantly to any server (see figure 2.2); memory streams to new virtual machine as needed and as transparently shared. These clone virtual machine start many times faster than traditional booted virtual machine and consume of fractions of memory and cutoff almost all IO booted.

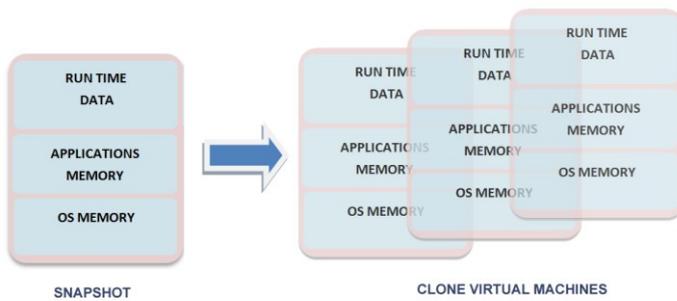

Figure 2.2: Snapshot

In the figure 2.3, through RAM cloning, running virtual machines are streamed instantly into new virtual machines.

Installing a guest operating system and applications can be time consuming. With clones, you can make many copies of a virtual machine from a single installation and configuration process. Cloning a virtual machine is faster than start it again. Users can launch many virtual machines in less than few seconds.

By this technique, VMS get ability to stream and thin provision the virtual machine's memory enables de-duplication of common elements for use by virtual servers. Virtual Memory Streaming (VMS) technique gives a true live migration to move VMs from one server to another without ever losing packets. VMS changes the sequence of tasks to eliminate the need to boot and configure virtual machines and the associated heavyweight data transfer.

### III. ARCHITECTURE OF PROPOSED MODEL AND IMPLEMENTATION DETAILS

This section describes the proposed architecture and implementation details of the solution developed to enable virtual memory streaming for VMs in order to automatically scaling of virtual infrastructure. In this work, we have focused on the open-source OpenStack [10] Open Source Cloud Computing Software, Piston OpenStack [11], KVM hypervisor [12], which have achieved wide use and is broadly accepted in the virtualization and open-source cloud community. Currently, the main open-source Virtual Infrastructure Managers (VIM) such as OpenNebula [15], OpenStack [10], CloudStack [16] and Eucalyptus [17] do not virtual memory streaming out of the box.

OpenStack extensions that enable VMS to run on OpenStack have been developed and made available the OpenStack Virtual Machine (VM) instances for both KVM and Xen Hypervisors [8] under the Apache license. After insatallation, OpenStack - A computer with the nova tool installed, as well as the VMS novaclient plugin. Therefore, for OpenStack, VMS is implemented as an OpenStack extension through Nova-API. A lightweight service that runs on each compute host and exposes an API primitive for nova called live-image-create. For this proof-of-concept implementation, we have modified OpenStack in order to include an additional operation to access the functionality of the KVM hypervisor for virtual memory streaming, through the Piston OpenStack. Therefore, the new operation allows to operation of live migration and virtual machine cloning from OpenStack. The figure 4.1 summarizes the main components for the model implementation to enable Virtual Memory Streaming (VMS) for virtual infrastructure.

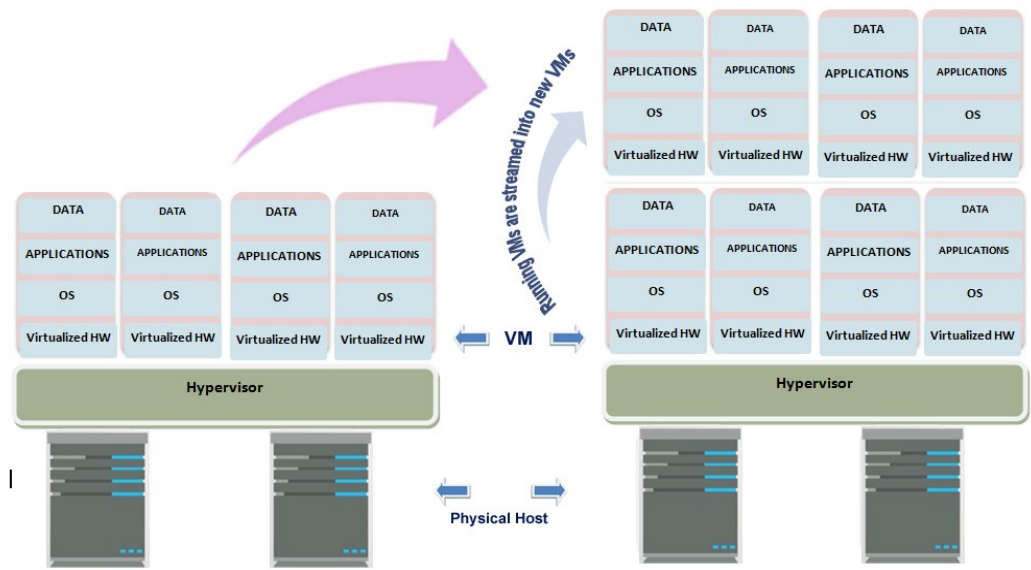

Figure: 2.3 running virtual machines are streamed instantly into new virtual machines.

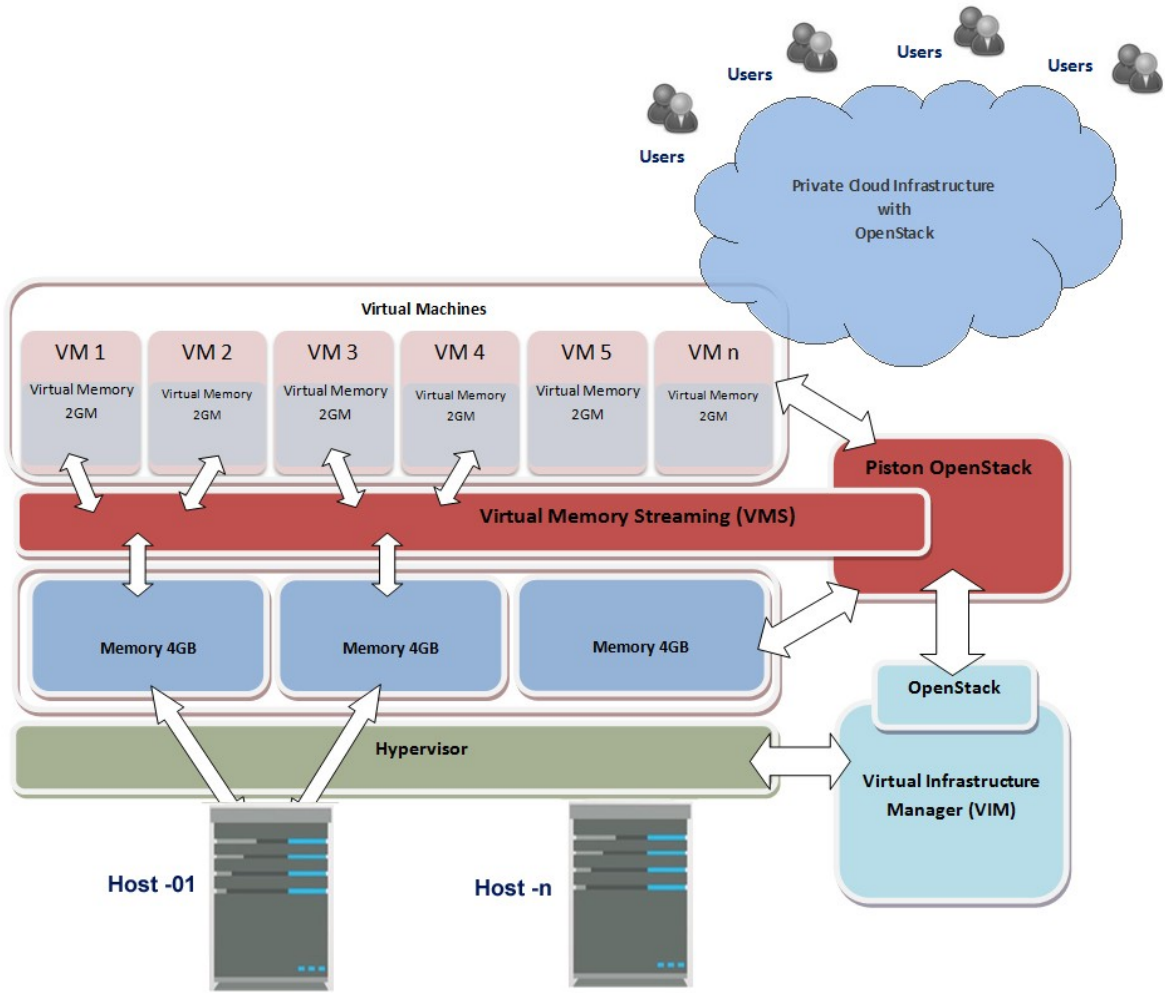

Figure 4.1: Model implementation to enable Virtual Memory Streaming (VMS) for virtual infrastructure.

First of all, the VMs are deployed via OpenStack on a private Cloud infrastructure. OpenStack is freely available under the Apache 2.0 license. They are based on a Virtual Machine Image (VMI) with Ubuntu 12.04 LTS, which has a kernel recent enough (3.2.X) to support memory streaming techniques when running as a guest OS. Now, we install Piston OpenStack. Piston OpenStack is a cloud operating system that allows to deploy and to manage an OpenStack private cloud without IT operations expertise. With VMS Piston OpenStack streams the memory from a instance snapshot of an actually booted memory that are running from VMs. Virtual machines launched through VMS start with a small amount of memory. Required memory is allocated for the VMs and optimized continuously. Through memory cloning, running virtual machines are streamed instantly into new virtual machines. First level of virtual machine provisioning- for each host a numbers of VMs provisioned with 4 GM virtual memory; after using VMS techniques – each VM is streamed into two VMs with 2 GB virtual memory. As virtual machines are thin provisioned, reduces memory footprints and thus, let pile on twice as many virtual servers onto every physical server. VMS techniques involve real-time policy-based memory management system that will control the memory footprint of each virtual machine as well as host-wide memory usage.

IV. CASE STUDY

In order to assess the effectiveness of these techniques, we have executed a case study on a private Cloud infrastructure composed by four Blade servers (M600 and M610 models). Each server has eight cores and 16 GB of RAM. VMs are deployed by OpenStack and Piston-OpenStack is employed to manage the memory streaming techniques supported by the KVM hypervisor. VMs are based on Ubuntu 12.04 LTS with 1

GB of (initial) memory size and 1 GB of swap space. VMs run on top of KVM which in turn runs on top of a host OS (which again is Ubuntu 12.04). One or more physical hosts running Ubuntu 12.04 LTS. A Havana, installation of OpenStack. These instructions assume Havana. Basic functionality like creation of images and booting of VMs is assumed to be working. Ubuntu Cloud Infrastructure is a ready to deploy Infrastructure-as-a-Service (IaaS) based on OpenStack. These instructions apply to Ubuntu 12.04 LTS to deploy Virtual Memory Streaming for OpenStack VMS version 2.6. After insatallation, OpenStack - A computer with the nova tool installed, as well as the VMS nova-client plugin. nova API, compute, and dashboard services can be running on different physical hosts or can all be running on the same host. These instructions assume that all services are running on the same host.

For Ubuntu, Havana installation of OpenStack. Basic functionality like creation of images and booting of VMs is assumed to be working. Install the API extension using the apt-get command. If you are installing to a Havana OpenStack installation on multi-node deployments, VMS requires shared storage to store live-image snapshots. VMS supports NFS, Ceph, and glance as a shared storage backend. These are the following steps for configuration for VMS: Install an Ubuntu image; Create a cloud-init script; Boot a VM - boot a VM as usual, but with the –user-data flag to pass the cloud-init file to the VM; Configure the VM -Once the VM has booted, we can install software and configure any post-launch steps; Create a VM snapshot - once the VM has been configured as desired, we can create a snapshot using the nova live-image-create command; We can check on the status of the snapshot using the nova list command; Launch a clone VM from snapshot - Once the snapshot is in the complete state, we can launch new VMs from it with the nova live-image-start command.

## V. EVALUATION

When a new virtual machine is launched, VMS streams the live image state to the new instance making it immediately available. The new instance starts up from where the snapshot left off. Common elements are de-duplicated and shared between virtual machines to reduce IO and storage needs. The new instance is automatically configured by VMS on the fly - when the new clone comes up, we change its network identity and hostname. VMS also seamlessly joins your new virtual machine to your active directory without a reboot.

VMS technique will do a rolling upgrade of an entire OpenStack environment without ever turning off a service or without ever turning off a VM. Traditional hypervisors such as KVM takes VM offline for anywhere between 15 and 30 seconds, which is not good enough. So Virtual Memory Streaming gives a true live migration to move VMs from one server to another without ever losing packets.

Some are the average VM startup time [17] showed in the table 5.1.

TABLE III.  AVERAGE VM STARTUP TIME

| Cloud | OS | Average VM startup time |
|---|---|---|
| EC2 | Linux | 96.9 seconds |
| EC2 | Windows | 810.2 seconds |
| Azure | WebRole | 374.8 seconds |
| Azure | WorkerRole | 406.2 seconds |
| Azure | VMRole | 356.6 seconds |
| Rackspace | Linux | 44.2 seconds |
| Rackspace | Windows | 429.2 seconds |

Table 5.1: average VM startup time from: source [17].

With VMS technique, in performance characteristics where handling 5% more load consumes very close to 5% more resources (CPU, memory, and I/O). This also allows for fine-grained packing of VMs into the available memory of physical hosts, putting all available memory to work without any drastic measures such as ballooning or re-sizing VMs. Instantiating machines this way allows the simultaneous launch of hundreds of instances instantaneously. And, in addition to doubling the virtual machine density, significantly cutting storage and network requirements and the budgets to back them. With VMS technique, user can double the virtual machines per host via memory oversubscription, thus cutting the required memory cost budget by half and spending that savings in scaling the infrastructure with other required resources. I/O bottlenecks are removed as user streams only 2-4GB of RAM which is faster than streaming 20-40GB of disk image. Using live images, VMS speeds up the time it takes to launch your virtual machines by 2x-10x. VMS dynamically allocates and streams memory on demand. This reduces the memory footprint of your virtual machines and increases the density of virtual machines per server by 2x-4x. As virtual machines are thin provisioned, reduces memory footprints and thus, let pile on twice as many virtual servers onto every physical server.

## VI. RELATED WORKS

Mao, Ming, and Marty Humphrey do a performance study on the VM startup time in the cloud [20]. C Clark, K Fraser, S Hand, JG Hansen, E Jul [21] demonstrate the migration of entire OS instances on a commodity cluster, recording service downtimes as low as 60ms. They show that that our performance is sufficient to make live migration a practical tool even for servers running interactive loads [21]. F Travostino, P Daspit, L Gommans, C Jog [22] demonstrate seamless live migration of virtual machines over the MAN/WAN. They work shows – "integration of VMs and deterministic light-path network services creates the VM Turntable demonstrator, an environment where in a VM is no longer limited to execute within the confines of a data center" [22]. H Liu, H Jin, X Liao, L Hu, C Yu [23] demonstrate live migration of virtual machines (VM) based on full system trace and replay. They shows live migration of virtual machines (VM) across distinct physical hosts provides a significant new benefit for administrators of data centers and clusters [23]. W Voorsluys, J Broberg, S Venugopal, R Buyya [24] shows capability of virtual machine live migration brings benefits such as improved performance, manageability and fault tolerance. R Bradford, E Kotsovinos, A Feldmann [25] present a design, implementation, and evaluation of a system for supporting the transparent, live wide-area migration of virtual machines that use local storage for their persistence state. MR Hines, K Gopalan [26] present a design and implementation of a post-copy technique for live migration of virtual machines. T Wood, PJ Shenoy, A Venkataramani, MS Yousif [27] argue that virtualization provides significant benefits in data centers by enabling virtual machine migration to eliminate hotspots. They present Sandpiper, a system that automates the task of monitoring and detecting hotspots, determining a new mapping of physical to virtual resources and initiating the necessary migrations in a virtualized data center. H Jin, L Deng, S Wu, X Shi [r8] argue that live migration of virtual machines has been a powerful tool to facilitate system maintenance, load balancing, fault tolerance, and power-saving, especially in clusters or data centers. They introduce memory compression technique into live VM migration [28]. Y Luo, B Zhang, X Wang, Z Wang [29] describe a whole-system live migration scheme, which transfers the whole system run-time state, including CPU state, memory data, and local disk storage, of the virtual machine (VM). They describe in the three phases - in the pre-copy phase, in the freeze-and-copy phase and in the post-copy phase [29].

Y Zhao, W Huang [30] describe a distributed load balancing algorithm based on live migration of virtual machines in cloud. K Ye, X Jiang, D Huang, J Chen [31] describe live migration framework of multiple virtual machines with resource reservation technology. They also perform a series of experiments to investigate the impacts of different resource reservation methods on the performance of live migration in both source machine and target machine [31]. H Liu, H Jin, X Liao, L Hu, C Yu [32] demonstrate live migration of virtual machine based on full system trace and replay. C Jo, E Gustafsson, J Son, B Egger [33] show live migration of virtual machines using shared storage. Y Ma, H Wang, J Dong, Y Li [34] show live migration of virtual machine with memory exploration and encoding. Two representative live migration systems, Xen live migration [18], and VMware VMotion [19], share similar implementation strategies. Both of them assume shared disk storage.

Virtual machine (VM) cloning is to create a replica of a source virtual machine (parent virtual machine); the replica, also called child virtual machine, owns exactly the same executing status as parent virtual machine. Fast live cloning guarantees that, during the period of cloning, the services running on the parent virtual machine observe no performance degradation [35], [39]. There are three important goals for fast live cloning: reducing the total cloning time, minimizing the suspension time of the parent virtual machine, and maximizing resource sharing between the parent virtual machine and the child virtual machine [39]. HA Lagar-Cavilla, JA Whitney, AM Scannell, R Bryant [36], [37] in their work they introduce VM fork and SnowFlock, and Xen-based implementation of virtual machine cloning with little runtime overhead, and frugal use of cloud IO resources. Y Sun, Y Luo, X Wang, Z Wang, B Zhang [39] exploit copy-on-write (CoW) mechanism to minimize the time of duplicating active main memory and secondary storage of the parent virtual machine. They argue that, the total cloning time of a virtual machine on Xen virtual machine monitor can be confined within several hundred milliseconds and the downtime of parent virtual machine is limited to tens of milliseconds, close to VMMpsilas scheduling interval [39]. MJ Mior, E de Lara [40] introduce FlurryDB, a scalable relational database with virtual machine cloning and argue that uses virtual machine cloning to improve resource utilization by drastically reducing the latency required to add a new replica. MJ Barber, DJ Ogden, A Aspinwall [41] demonstrate clone multiple different types of operating system images within a Virtual Machine.

Through, so many these researchers work on their very special field in live migration, cloning, caching, page-sharing or footprint management; they do not give a technique with combination some of these for efficient virtual machine scaling.

VII. CONCLUSION

Virtual Memory Streaming (VMS) is a scaling virtualization technology that allows different virtual machines rapid scale, high performance, and increase hardware utilization. VMS brings together several advanced hypervisor memory management techniques including granular page sharing, dynamic memory footprint management, live migration, cloning, read caching, and a unique virtual machine cloning capability.

An architecture model is described, together with a proof-of-concept implementation, that VMS dynamically scaling of virtualized infrastructure with true live migration and cloning of VMs. VMS provide rapid horizontal scale - instantly adding more virtual machines and growing from one to ten of virtual machines within few seconds. Scale interactive workloads on demand by scaling the number of virtual machines up or down automatically based on resource usage. Provide high performance – eliminate booting of virtual machines by launching VMs quickly from live images within a short time. This technique also reduces memory footprints as virtual machines are thin provisioned and memory is streamed on demand. VMS allows operators of both public and private clouds to scale their virtual infrastructures automatically and instantaneously while substantially reducing memory requirements. By using Virtual Memory Streaming (VMS) technique increased hardware utilization; as well as decreased hardware and software capital costs and thus decreased operating costs.

REFERENCES

[1] Denning, Peter J. "Virtual memory." ACM Computing Surveys (CSUR) 2.3 (1970): 153-189.

[2] Bugnion, Edouard, Scott W. Devine, and Mendel Rosenblum. "Virtual machine monitors for scalable multiprocessors." U.S. Patent No. 6,075,938. 13 Jun. 2000.

[3] Barham, Paul, et al. "Xen and the art of virtualization." ACM SIGOPS Operating Systems Review 37.5 (2003): 164-177.

[4] Goldberg, Robert P. "Survey of virtual machine research." Computer 7.6 (1974): 34-45.

[5] Bugnion, Edouard, Scott W. Devine, and Mendel Rosenblum. "Virtual machine monitors for scalable multiprocessors." U.S. Patent No. 6,075,938. 13 Jun. 2000.

[6] Rosenblum, Mendel, and Tal Garfinkel. "Virtual machine monitors: Current technology and future trends." Computer 38.5 (2005):39-47.

[7] Armbrust, Michael, et al. "A view of cloud computing." Communications of the ACM 53.4 (2010): 50-58.

[8] From web:.xenproject.org/

[9] Mao, Ming, and Marty Humphrey. "A performance study on the vm startup time in the cloud." Cloud Computing (CLOUD), 2012 IEEE 5th International Conference on. IEEE, 2012.

[10] From web:openstack.org/

[11] From web:pistoncloud.com/openstack-cloud-software/

[12] From web:linux-kvm.org/page/Main_Page

[13] From web:opennebula.org/

[14] From web:cloudstack.apache.org/

[15] From web:www.eucalyptus.com/

[16] Kivity, Avi, et al. "kvm: the Linux virtual machine monitor." Proceedings of the Linux Symposium. Vol. 1. 2007.

[17] Mao, Ming, and Marty Humphrey. "A performance study on the vm startup time in the cloud." Cloud Computing (CLOUD), 2012 IEEE 5th International Conference on. IEEE, 2012.

[18] From web wiki.xen.org/wiki/Storage_XenMotion

[19] From web:vmware.com/products/vsphere/features/vmotion.html

[20] Mao, Ming, and Marty Humphrey. "A performance study on the vm startup time in the cloud." Cloud Computing (CLOUD), 2012 IEEE 5th International Conference on. IEEE, 2012.

[21] Clark, Christopher, et al. "Live migration of virtual machines." Proceedings of the 2nd conference on Symposium on Networked Systems Design & Implementation-Volume 2. USENIX Association, 2005.

[22] Travostino, Franco, et al. "Seamless live migration of virtual machines over the MAN/WAN." Future Generation Computer Systems 22.8 (2006): 901-907.

[23] Liu, Haikun, et al. "Live migration of virtual machine based on full system trace and replay." Proceedings of the 18th ACM international symposium on High performance distributed computing. ACM, 2009.

[24] Voorsluys, William, et al. "Cost of virtual machine live migration in clouds: A performance evaluation." Cloud Computing. Springer Berlin Heidelberg, 2009. 254-265.

[25] Bradford, Robert, et al. "Live wide-area migration of virtual machines including local persistent state." Proceedings of the 3rd international conference on Virtual execution environments. ACM, 2007

[26] Hines, Michael R., and Kartik Gopalan. "Post-copy based live virtual machine migration using adaptive pre-paging and dynamic self-ballooning." Proceedings of the 2009 ACM SIGPLAN/SIGOPS international conference on Virtual execution environments. ACM, 2009.

[27] Wood, Timothy, et al. "Black-box and Gray-box Strategies for Virtual Machine Migration." NSDI. Vol. 7. 2007.

[28] Jin, Hai, et al. "Live virtual machine migration with adaptive, memory compression." Cluster Computing and Workshops, 2009. CLUSTER'09. IEEE International Conference on. IEEE, 2009.

[29] Luo, Yingwei, et al. "Live and incremental whole-system migration of virtual machines using block-bitmap." Cluster Computing, 2008 IEEE International Conference on. IEEE, 2008.

[30] Travostino, F., Daspit, P., Gommans, L., Jog, C., De Laat, C., Mambretti, J., ... & Yonghui Wang, P. (2006). Seamless live migration of virtual machines over the MAN/WAN. Future Generation Computer Systems, 22(8), 901-907.

[31] Ye, Kejiang, et al. "Live migration of multiple virtual machines with resource reservation in cloud computing environments." Cloud Computing (CLOUD), 2011 IEEE International Conference on. IEEE, 2011.

[32] Liu, Haikun, et al. "Live migration of virtual machine based on full system trace and replay." Proceedings of the 18th ACM international symposium on High performance distributed computing. ACM, 2009.

[33] Jo, Changyeon, et al. "Efficient live migration of virtual machines using shared storage." Proceedings of the 9th ACM SIGPLAN/SIGOPS international conference on Virtual execution environments. ACM, 2013.

[34] Ma, Yanqing, et al. "ME2: Efficient Live Migration of Virtual Machine with Memory Exploration and Encoding." Cluster Computing (CLUSTER), 2012 IEEE International Conference on. IEEE, 2012.

[35] Hutchins, Greg, et al. "Offloading operations to a replicate virtual machine." U.S. Patent No. 8,296,759. 23 Oct. 2012.

[36] Lagar-Cavilla, Horacio Andrés, et al. "SnowFlock: rapid virtual machine cloning for cloud computing." Proceedings of the 4th ACM European conference on Computer systems. ACM, 2009.

[37] Lagar-Cavilla, H. Andrés, et al. "Snowflock: Virtual machine cloning as a first-class cloud primitive." ACM Transactions on Computer Systems (TOCS) 29.1 (2011): 2.

[38] Chen, Peter M., and Brian D. Noble. "When virtual is better than real [operating system relocation to virtual machines]." Hot Topics in Operating Systems, 2001. Proceedings of the Eighth Workshop on. IEEE, 2001.

[39] Sun, Yifeng, et al. "Fast live cloning of virtual machine based on xen." High Performance Computing and Communications, 2009. HPCC'09. 11th IEEE International Conference on. IEEE, 2009.

[40] Mior, Michael J., and Eyal de Lara. "Flurrydb: a dynamically scalable relational database with virtual machine cloning." Proceedings of the 4th Annual International Conference on Systems and Storage. ACM, 2011.

[41] Barber, Michael J., et al. "Creation of temporary virtual machine clones of multiple operating systems." U.S. Patent Application 11/589,707.